\documentstyle[preprint,aps]{revtex}

\tighten

\begin{document}

\preprint{WM-94-110}

\draft

\title{Non-factorization and the Decays $B \rightarrow J/\psi+K^{(*)}$}

\author{Carl E. Carlson}
\address{Physics Department, College of William and Mary,
Williamsburg, VA 23187}

\author{J. Milana}
\address{Physics Department, University of Maryland,
College Park,  MD  20742}

\date\today

\maketitle

\vglue-0.3in

\begin{abstract}

Many known models, which generally use a
factorization hypothesis,  give a poor account of the decays $B
\rightarrow J/\psi + K^{(*)}$.  Usually there is a free overall
factor, which is fit to the data, so that tests of the models rely
upon ratios.  The models tend to give too much $K^*$ compared to
$K$ and too much transverse polarization compared to longitudinal.
Our microscopic calculations, which use perturbative QCD, do
well for both ratios. A microscopic calculation allows us to see
how well factorization, heavy quark symmetry, and other features of
various models are working. In the present case,
agreement with the experimental ratios is dependent upon a
breakdown of factorization for one of the amplitudes.
\end{abstract}

\pacs{PACS numbers: 13.20.He, 12.15.-y, 12.38.Bx}

\vglue0.25in

\section{Factorization and data}

Gourdin, Kamal, and Pham~\cite{gkp94} and Aleksan {\it et
al.}~\cite{alopr94} point out that many known
models~\cite{bsw87,casal92,ali91,jaus90},  all of which use the
factorization hypothesis,  give a poor account of the decays $B
\rightarrow J/\psi + K^{(*)}$.

In most models there is an overall factor, generally called
$a_2$~\cite{bsw87}, which is fit to the data, so that tests of the
models rely upon ratios of $K^*$ and $K$ decays, and ratios of
longitudinal and tranverse polarization in the $J/\psi+K^*$ decays.
The models tend to give too much $K^*$ compared to $K$ and too much
transverse polarization compared to longitudinal.

Our microscopic calculations, which use perturbative QCD, do well for
both ratios. Although the results have been published in some
detail~\cite{cm94,cm93,shb90}, the charmonium $B$ decays deserve some
further thought because of the present interest in them, and we will
attempt to make self-contained at least the qualitative parts of our
present remarks. A microscopic calculation allows us to see how well
factorization, heavy quark symmetry, and other features of various
models are working. In the present case, we find a serious breakdown of
factorization for one of the amplitudes.

More explicitly, the ratios under study and
their experimental values are~\cite{cleo94},
\begin{equation}
R\equiv{{\rm Br}(B\rightarrow J/\psi + K^*)\over
       {\rm Br}(B\rightarrow J/\psi + K)  }
   = 1.71 \pm 0.40,
\end{equation}
and, dividing the $K^*$ rate into a longitudinal polarization
part $\Gamma_{L}$ and a transverse one $\Gamma_{T}$,
\begin{equation}
\left(\Gamma_{L}\over \Gamma\right)_{K^*} = \left\{
\begin{array}{ll}
0.80\pm0.08\pm0.05 & {\rm \quad CLEO~\cite{cleo94}} \\
0.66\pm0.10\pm0.10 & {\rm \quad CDF~\cite{cdf94}}
\end{array}
\right.
\end{equation}
Our own results for the two ratios are 1.76 and 0.65, respectively
(using Table IV of~\cite{cm94}).

Factorization implies that the decays depend upon a set of form
factors for a current connecting $B$ to $K^{(*)}$.  As a
benchmark---yes, we know the $K^{(*)}$ is light---the relations that
heavy quark symmetry~\cite{nw87,vs88,iw89} implies among the form
factors lead to
\begin{equation}
R={m_B^2+4m_{J/\psi}^2 \over m_B^2} \approx 2.38
\end{equation}
and
\begin{equation}
\left(\Gamma_{L}\over \Gamma\right)_{K^*} =
       {m_B^2 \over m_B^2+4m_{J/\psi}^2} \approx 0.42,
\end{equation}
which, although they do not agree with the data, are not bad
as a representation of many of the models.  For information, in each
combination $m_B^2+4m_{J/\psi}^2$, the $m_B^2$ comes from
$\Gamma_{L}$ and the $4m_{J/\psi}^2$ comes from $\Gamma_{T}$.

Why does our calculation work for the ratios when others do not? Most
importantly, the factorization hypothesis fails. It does not fail
uniformly. Its failure is significant only for the transverse
polarization final state of $B\rightarrow J/\psi + K^*$.  In this
amplitude the nonfactorizable contributions are about half the size
and opposite in sign to the factorizable ones, which has roughly the
effect of turning the ``4'' into a ``1'' in the previous equations,
and giving decent agreement with the $\left(\Gamma_{LL} /
\Gamma\right)_{K^*}$ data.

Also, surprisingly in this context, we find the
heavy quark symmetry symmetry predictions for the form factors of the
factorizable parts of the amplitude work surprisingly well. One might
expect significant differences due to a nonperturbative cause, namely
that the wave functions or distribution amplitudes of the $K$ and $K^*$
are different.   A wave function difference at the origin is shown by
data that gives unequal decay constants for the $K$ and $K^*$, and the
shapes of the two wave functions are also different. We use the
distribution amplitudes for $K$ and $K^*$ worked out from QCD sum rules
by Chernyak, Zhitnisky, and Zhitnitsky~\cite{czz82}.  The upshot is that
the form factors relative to the heavy quark symmetry predictions are
good, and that small corrections and nonfactorizable contributions keep
the two ratios from just being inverses of each other.

Some details will now come.

\section{More detailed discussion}

\subsection*{Factorization}

We should state what factorization means in the context of $B
\rightarrow J/\psi+K^{(*)}$.  The relevant part of the effective
Hamiltonian density is
\begin{equation}
H_{eff} = {G_F\over\sqrt{2}}V_{cb}V_{cs}\
           \bar s \gamma_{\mu L} c  \
           \bar c \gamma^\mu_L b ,
\end{equation}
where $\gamma^\mu_L = \gamma^\mu (1-\gamma_5)$ and the matrix element
we want is, generically,
\begin{equation}
M = \langle X,\psi | H_{eff} | B \rangle  ,
\end{equation}
 where $\psi= \eta_c, J/\psi, \ldots$.

The factorization hypothesis is that the charmed quarks which
are created go into the $\psi$ and, except for the weak interaction
vertices, are unconnected to other quarks in the process.  We also
assume that the outgoing charmed quarks in the $\psi$ have
small transverse momentum relative to the direction of the $\psi$.
If the factorization hypothesis is valid, one can show
\begin{equation}
M =-{1\over N_c} {G_F\over\sqrt{2}}V_{cb}V_{cs}\
     (\bar c \gamma_{\mu L} c)_\psi \
     (\bar s \gamma^\mu_L b)_{B\rightarrow X},
\end{equation}
that is, the matrix element is a product of two hadronic factors,
\begin{equation}
(\bar c \gamma_{\mu L} c)_\psi \equiv
    \langle \psi | \bar c \gamma_{\mu L} c | 0 \rangle
\end{equation}
and
\begin{equation}
(\bar s \gamma^\mu_L b)_{B\rightarrow X} \equiv
    \langle X | \bar s \gamma^\mu_L b | B \rangle  .
\end{equation}
Quantity $N_c$ is the number of colors, and to allow for effects
of mixing with other operators one usually replaces $(-1/N_c)$ with
a constant $a_2$ which is determined by data.

\subsection*{Non-factorization}

In Fig. \ref{BJdecay}, parts (a) and (d) correspond to the
factorizable contributions, in the present context, and parts (b) and
(c) to the nonfactorizable ones.

Upon first view, it is easy to believe that the nonfactorizable
contributions are small. The gluon couples to two oppositely colored
quarks that are nearly at the same point because of the
$W$-exchange.  Indeed, the largest parts of diagrams (b) and (c)
cancel each other and the subleading $O(q_G)$, where $q_G$ is
the gluon momentum, terms give the surviving result. The pieces of
Figs.~\ref{BJdecay} (b) and (c) from one weak vertex, through the
$J/\psi$ (whose polarization vector is $\xi$) including the gluon
emmission vertex ($\gamma^\nu$), and to the other weak vertex, have
numerators that sum to
\begin{equation}
4m_{J/\psi} (1+\gamma_5)
   \left(
   {\not}\kern2pt\xi \,\gamma^\nu {\not}\kern1.75ptq_G -
          {\not}\kern1.75ptq_G \,\gamma^\nu {\not}\kern2pt\xi
   \right)
   (1-\gamma_5)  ,
\end{equation}
which does go to zero for gluons of long
wavelength, or $q_G$ going to zero.  (The
numerators of Figs.~\ref{BJdecay} (a) and (d) do not go to
zero in the same limit.)  However, the gluon momentum is not so small;
in fact we argue that it is large enough that a perturbative
calculation is plausibly valid. It supplies the momentum transfer
needed by the light quark, which is of order
$\bar\Lambda_B$, the part of the mass of the $B$ meson carried by the
light quark, which is about 500 Me{\kern-1.0pt}V or a few times
$\Lambda_{QCD}$.

However, for $B \rightarrow J/\psi + K$ and the longitudinal part of
$B\rightarrow J/\psi +K^*$, there is further cancellation between the
subleading parts of the two nonfactorizable diagrams.  In contrast,
they add for the transverse decay, so this nonfactorizable amplitude
can get large.  While the transverse $B
\rightarrow J/\psi + K^*$ does require chirality violations, the
ensuing suppression is of
$O(m_{J/\psi}/m_B)$, which is not a decisive factor.

\section{Closing Remarks}

We have seen, in one explicit calculation, how nonfactorizable
contributions to $B \rightarrow J/\psi + K^{(*)}$ are significant and
are  crucial to giving agreement with data for the $K^*$ to $K$ ratio
and the tranverse to longitudinal polarization ratio.

Ratios are used to test the models because in most models the overall
rate is determined by a constant that is fit to the data.
Our calculations using perturbative QCD also have trouble with the
overall rate.  At present, with the parameters we choose~\cite{cm94}, we
do a good job on non-color-suppressed decays such as $B \rightarrow D
\pi$ but the overall rates are rather low compared to data for decays
like $B
\rightarrow J/\psi + K^{(*)}$.

We believe the use of pQCD is valid for high recoil decays of the
$B$.  This has been given better support by Akhoury, Sterman, and
Yao~\cite{asy94}who consider Sudakov effects in pQCD calculations of
$B$ decays and show that they suppress contributions from end point
regions where use of pQCD would be questionable.

It is possible that perturbative contributions are part but not all of
what gives $B \rightarrow J/\psi + K^{(*)}$ decay. In this case, the
details of our remarks will only be part of something larger, but the
scenario can well stand:  the factorizable contributions to  $B
\rightarrow J/\psi + K^{(*)}$ will not suffice to explain those
decays, and nonfactorizable contributions will be crucial.

%\newpage

\appendix

\section{Factorization hypothesis}

Here is a perturbative proof that the factorization hypothesis leads
to the factored form of the amplitude. Take the
$J/\psi$ as an example. Factorization allows us to isolate the $\psi$
piece,
\begin{eqnarray}
&&\bar s_i \gamma_\mu(1-\gamma_5)
   \langle J/\psi | c_i \bar c_j | 0 \rangle
   \gamma^\mu(1-\gamma_5) b_j \nonumber \\
&&= \bar s_i \gamma_\mu(1-\gamma_5)
  \left[ {f_{J/\psi} \over 2\sqrt{N_c}}
  {\delta_{ij}\over \sqrt N_c}
  {\not{\kern-1pt \xi} (\not{\kern-2pt q}+m_{J/\psi})\over \sqrt 2}
                                                             \right]
  \gamma^\mu(1-\gamma_5) b_j \nonumber \\
&&= -{1\over N_c} \left[ \sqrt{2} f_{J/\psi} m_{J/\psi} \xi_\alpha
                                   \right]
  \bar s \gamma^\alpha_L b,
\end{eqnarray}
where $q$ and $\xi$ are the momentum and polarization vectors of
the $J/\psi$ and $i$ and $j$ are color indices. One needs to
recognize
\begin{equation}
\langle J/\psi | \bar c \gamma_\alpha c | 0 \rangle
= \sqrt{2} f_{J/\psi} m_{J/\psi} \xi_\alpha
\end{equation}
to complete the proof.

\section{Factorizable amplitudes}

It is of some interest to see the behavior of the factorizable parts
of the amplitudes.  In practice, diagram \ref{BJdecay}(d) is quite
small and we will give the results from Fig. \ref{BJdecay}(a).  We use
form factors $f_\pm$, $a_\pm$, $g$, and $f$ as defined in
ref.~\cite{iw89}. Neither $f_-$ nor $a_-$ enters when the decay involves
the $J/\psi$. If desired, one may convert to ref.~\cite{bsw87}
definitions by ($q^2
\leftrightarrow m_{J/\psi}^2$ here),
\begin{eqnarray}
F_1 &=& f_+ \nonumber \\
F_0 &=& {q^2\over m_B^2-m_K^2}f_- +f_+ \nonumber \\
V   &=& -(m_B+m_{K^*}) g \nonumber \\
A_1 &=& (m_B+m_{K^*})^{-1} f \nonumber \\
A_2 &=& -(m_B+m_{K^*}) a_+ \nonumber \\
A_0 &=& (2m_{K^*})^{-1} (f+(m_B^2-m_{K^*}^2)a_+ +q^2 a_-)
\end{eqnarray}
We keep the explicit $m_{K^{(*)}}$ mass terms although they turn
out to have small effect (unless necessary in a definition).

Each form factor can be written like
\begin{equation}
f_+ = B\int_0^1 dy_1\, \tilde\phi_K {(1-y_1) (a+by_1)
      \over y_1 - r -i\eta },
\end{equation}
where
\begin{equation}
B = {16\pi \alpha_s f_B f_{K^{(*)}} \over
    3 \epsilon_B (m_B^2 +m_{K^{(*)}}^2 - m_{J/\psi}^2)^2 }
\end{equation}
for $\epsilon_B = \bar\Lambda_B/m_B$ and
\begin{equation}
r = { 2 \epsilon_B m_B^2 \over m_B^2 +m_{K^{(*)}}^2 - m_{J/\psi}^2}.
\end{equation}
Quantity $\tilde\phi_K$ is the distribution amplitude of the kaon with
the asyptotic form factored out (so that $\tilde\phi_K = 1$ if we wish
to use the asymptotic form).  For form factor $f_+$,
\begin{eqnarray}
a &=& a(f_+) = m_B(m_B+m_K) - \epsilon_B m_B(m_B-m_K) \nonumber \\
b &=& b(f_+) = m_B^2 - 2m_Bm_K - m_{J/\psi}^2.
\end{eqnarray}
For the other form factors,
\begin{eqnarray}
a(g) &=& m_B(1+ \epsilon_B) \nonumber \\
b(g) &=& -m_{K^*}
\end{eqnarray}
and
\begin{eqnarray}
a(f) &=& 2m_B^2m_{K^*}(1 - 2\epsilon_B) \nonumber \\
  &&+ (m_B^2 +m_{K^*}^2 - m_{J/\psi}^2) m_B (1+\epsilon_B) \nonumber\\
b(f) &=& m_{K^*} (m_B^2 +m_{K^*}^2 -4m_Bm_{K^*}- m_{J/\psi}^2).
\end{eqnarray}
It happens that the heavy quark symmetry results $a_\pm =\pm g$ are
obeyed exactly.

The integrals can be done analytically, using
\begin{eqnarray}
I_N &=& \int dy\, {y^N \over y-r-i\eta} =
   \sum_{k=0}^{N-1} {1\over N-k}r^k + r^N I_0,
\end{eqnarray}
for $N \ge 1$ and
\begin{equation}
I_0 = i\pi + \ln \left(1-r\over r \right) .
\end{equation}

For asymptotic wave functions
\begin{eqnarray}
{|f_+|\over m_B+m_K}:{|g|}&:&
{|f|\over (m_B-m_{K^*})^2-m_{J/\psi}^2}  \nonumber \\
 = 0.99 f_K&:&1.03 f_{K^*}:0.99 f_{K^*}.  \label{one}
\end{eqnarray}
Heavy quark symmetry predicts 1:1:1.  Of course, strict heavy
quark symmetry also predicts $f_{K^*}/f_K = 1$ whereas the experimental
result is $1.67f_\pi/ 1.22f_\pi \approx 1.37$.

One should perhaps not use the asymptotic distribution amplitudes for
the kaons. A common form for representing the distribution amplitude is
\begin{equation}
\tilde\phi(y_1) = 5\beta(2y_1-1)^2 + (1-\beta),
\end{equation}
where $y_1$ is the momentum fraction carried by the nonstrange quark
and $\beta$ is the fraction of the distribution amplitude that is not
asymptotic. The QCD sum rule results of Chernyak, Zhitnitsky, and
Zhitnitsky, lead to $\beta = 0.6$ for the $K$ and $0.1$ for the
$K^*$.  One then gets
\begin{equation}
1.38 f_K:1.03 f_{K^*}:1.03 f_{K^*}
\end{equation}
for the same ratio as eqn.~(\ref{one}) (with the same overall
constants).  This is stunningly close to the heavy quark symmetry
result!

\begin{figure}
\vglue 2.4in
\hskip 1in {\special{picture BJdecay scaled 1000}} \hfil
\vglue 0.1in
\caption{Lowest order perturbation theory diagrams for $B$ decays
involving charmonium. (a) and (d) are factorizable, (b) and (c) are
nonfactorizable.}
  \label{BJdecay}

\end{figure}

\end{document}